\documentclass[10pt,preprint]{aastex}
\begin{document}

\title{THE PURPLE HAZE OF ETA CARINAE: BINARY-INDUCED VARIABILITY?\altaffilmark{1}}

\author{Nathan Smith\altaffilmark{2,3}, Jon A.\ Morse\altaffilmark{4},
Nicholas R.\ Collins\altaffilmark{5,6}, and Theodore R.\
Gull\altaffilmark{5}}

\altaffiltext{1}{Based on observations made with the NASA/ESA {\it
Hubble Space Telescope}, obtained at the Space Telescope Science
Institute, which is operated by the Association of Universities for
Research in Astronomy, Inc., under NASA contract NAS5-26555.}

\altaffiltext{2}{Hubble Fellow}

\altaffiltext{3}{Center for Astrophysics and Space Astronomy, University of
Colorado, 389 UCB, Boulder, CO 80309}

\altaffiltext{4}{Department of Physics and Astronomy, Arizona State University,
Box 871504, Tempe, AZ 85287-1504}

\altaffiltext{5}{Laboratory for Astronomy and Solar Physics, NASA Goddard Space
Flight Center, Code 681, Greenbelt, MD 20771}

\altaffiltext{6}{Science Systems and Applications, Inc., 10210
Greenbelt Rd.\ Suite 600, Lanham, MD 20706}

\begin{abstract}

Asymmetric variability in ultraviolet images of the Homunculus
obtained with the Advanced Camera for Surveys/High Resolution Camera
on the {\it Hubble Space Telescope} suggests that $\eta$ Carinae is
indeed a binary system.  Images obtained before, during, and after the
recent ``spectroscopic event'' in 2003.5 show alternating patterns of
bright spots and shadows on opposite sides of the star before and
after the event, providing a strong geometric argument for an
azimuthally-evolving, asymmetric UV radiation field as one might
predict in some binary models. The simplest interpretation of these
UV images, where excess UV escapes from the secondary star in the
direction away from the primary, places the major axis of the
eccentric orbit roughly perpendicular to our line of sight, sharing
the same equatorial plane as the Homunculus, and with apastron for the
hot secondary star oriented toward the southwest of the primary.
However, other orbital orientations may be allowed with more
complicated geometries.  Selective UV illumination of the wind and
ejecta may be partly responsible for line profile variations seen in
spectra.  The brightness asymmetries cannot be explained plausibly
with delays due to light travel time alone, so a single-star model
would require a seriously asymmetric shell ejection.

\end{abstract}


\section{INTRODUCTION}

Since $\eta$~Carinae is potentially the most massive, most luminous,
and most unstable star known, it would be valuable to know if it is
indeed an interacting binary system.  A debate concerning the
possibility that $\eta$~Car may be a binary (or more specifically,
what role a companion star plays) has been ongoing since a 5.5 yr period was discovered
by Damineli (1996; see also Corcoran et al.\ 2001; Damineli et al.\
1997, 2000; Davidson 1999; Davidson et al.\ 2000; Duncan \& White
2003; Duncan et al.\ 1997; Ishibashi et al.\ 1999; Whitelock et al.\
2004; as well as contributions in Morse et al.\ 1999).  Part
of the reason for the ongoing debate is that spectroscopic changes
seem to be explained equally-well with either a binary system or a
shell ejection (Zanella et al.\ 1984; Davidson 1999; Smith et al.\
2003a).  Furthermore, spectroscopy of the central star with the {\it
Hubble Space Telescope} ({\it HST}) by Davidson et al.\ (2000) did not
confirm Doppler shifts measured in ground-based data that were
interpreted as orbital reflex motion (Damineli et al.\ 1997).
However, continued monitoring in X-rays and the near-infrared (IR) has
established a period of $\sim$2023 days (Damineli et al.\ 2000;
Whitelock et al.\ 2004), and the very hard X-ray spectrum is difficult
to explain without a colliding-wind binary (Corcoran et al.\ 2000,
2001; Ishibashi 2001; Ishibashi et al.\ 1999; Pittard \& Corcoran
2002).  If $\eta$~Car is a binary, the most plausible scenario
involves a very massive primary with a dense wind, and a less massive
($\sim$30 M$_{\odot}$) hot O-type secondary with a 2,000--3,000 km
s$^{-1}$ wind in a highly-eccentric orbit (Corcoran et al.\ 2001;
Ishibashi 2001; Pittard \& Corcoran 2002; Morse et al.\ 1999).

Yet, this hypothetical companion star continues to evade direct
detection as its emission is dwarfed at most wavelengths by the
primary.  Since the primary star's dense wind extinguishes much of its
own far-UV luminosity (Hillier et al.\ 2001), and because the
hypothetical companion star is probably an O-type star, we may have
the best hope of detecting the companion's radiation in the UV.

The ``Purple Haze'' is a diffuse blueish/purple glow within a few
arcseconds of the central star in {\it HST} images of the Homunculus
(Morse et al.\ 1998; Smith et al.\ 2000, 2004).  This emission is seen
in excess of violet starlight scattered by dust, and the strength of
the excess increases into the far UV (Smith et al.\ 2004; hereafter
Paper I).  This increasing excess is UV emission arising in the outer
parts of $\eta$ Car's stellar wind, even if it is a single star (Paper
I; Hillier et al.\ in prep.).  In Paper I we argued that the UV excess
emission originates near the equatorial plane, confined by a dust
torus seen at thermal-IR wavelengths (Smith et al.\ 1998, 2002, 2003b;
Morris et al.\ 1999), although there may also be a contribution from
the Little Homunculus nebula (Ishibashi et al.\ 2003).  This
equatorial UV emission probably results from a latitude-dependent
radiation field, as UV escapes more easily from lower-density
equatorial zones in the bipolar wind of $\eta$ Car (Smith et al.\
2003a).  Note that the wind of $\eta$ Car is latitude dependent,
regardless of whether that wind structure is caused by a rotating
primary, induced by a companion star, or both (Smith et al. 2003a).

If a hot companion star contributes to this UV excess, then
observations of the ``Purple Haze'' may offer a critical geometric
test of the binary hypothesis.  Specifically, if $\eta$ Car has a hot
companion, its UV radiation should escape preferentially in directions
away from the primary, with a UV ``shadow'' cast on the opposite side
of the primary near periastron, when the primary's dense wind would
obstruct the secondary's far-UV light over a substantial and
time-varying solid angle.  This assumption provides a simple but
strong qualitative prediction, and here we present evidence that this
binary scenario is applicable to $\eta$ Car.

\section{OBSERVATIONS}

To study potential changes in the Purple Haze, we used UV and
visual-wavelength images of $\eta$~Car from the Advanced Camera for
Surveys/High Resolution Camera (ACS/HRC) on {\it HST} using the F220W,
F250W, F330W, and F550M filters obtained as part of the $\eta$ Car
Treasury program (P.I.: Davidson).  The various contributions of
continuum and emission lines in each filter have been described in
Paper I.  All four filters were employed on several occasions before,
during, and after the mid-2003 spectroscopic event, as summarized in
Table 1.  After the initial observations in 2002 October, we designed
the remaining observations so that the two field stars located
30$\arcsec$ northwest of the central star would always be included in
the HRC field of view, in order to facilitate accurate spatial
alignment of the various epochs.  Because of limited space here, we
refer the reader to more detailed discussions of our data reduction
procedures (including careful corrections for geometric distortion)
described in Paper I, Morse et al.\ (1998, 2001), and Smith et al.\
(2000).

The multi-epoch images showed subtle intensity changes from one epoch
to the next, which are, however, difficult to convey on the printed
page.  To reveal this flux variability as a function of position in
the nebula, Figure 1 shows difference images --- i.e. each frame of
Figure 1 shows the result of the original image at the indicated date
with the average of all epochs subtracted.  Thus, bright knots in
Figure 1 are brighter than the 1 yr average, and dark regions are
fainter than average.  The intensity range in Figure 1 is
$\pm$10$^{-12}$ ergs s$^{-1}$ cm$^{-2}$ \AA$^{-1}$ arcsec$^{-2}$,
which is very roughly $\pm$25\% of the average flux at about
0$\farcs$5 from the central star where the most pronounced variability
is seen, and is roughly a factor of 10$^3$ larger than the 1$\sigma$
subtraction residuals near the edges of the images.  Figure 1 shows
results for the F220W filter, where the variability is the most
dramatic.  The F250W and F330W filters show identical qualitative
changes (confirming that these are not artifacts due to errors in
spatial alignment, polarization losses, or other effects), but the
variability is less pronounced at longer wavelengths as the Purple
Haze diminishes (compared to F220W, the excess is roughly 80\% and
30\% as strong in the F250W and F330W filters, respectively; Paper I).

Table 1 also lists the phase of each observation in the 5.5 yr cycle.
These were calculated assuming that $\phi$=0 is JD=2452840.3, defined
by the minimum of the K-band light curve reported by Whitelock et al.\
(2004), and adopting their period of 2023 days.  However, for extended
features there is an effective phase delay due to light travel time.
Most of the variability we describe below is typically at
0$\farcs$5--1$\arcsec$ from the star.  Delay times are a couple of
weeks to a month, requiring a correction as large as
$\Delta\phi$=--0.01 to --0.02.  With that correction, the 2003 Sep
image is closest to the actual event as seen by circumstellar ejecta,
and the 2003 Jul image traces ejecta that see the pre-event radiation
field.  Another effect near periastron, simlar to a time delay, is
that UV radiation from the secondary star will escape the primary's
wind through a ``cavity'' created by its own less dense stellar wind
which trails behind the secondary in its orbit.

\section{DISCUSSION}

\subsection{Variability of the Purple Haze}

The UV variability in Figure 1 is a relatively small perturbation on
the observed brightness distribution of the Homunculus.  It represents
roughly 25\% of the average total flux at a given position, and only
about 1/3 of the excess UV emission at 0$\farcs$5 from the star in the
2002 Oct F220W image (see Paper I).  Thus, the UV excess is still
dominated by the intrinsic UV radiation-field geometry of the primary
star, which is expected to have a strong 2200 \AA\ excess at large
radii in the wind (Paper I).  In other words, the 2200 \AA\ morphology
at all epochs during the event looks basically the same as described
in Paper I.  \footnotemark\footnotetext{Caveat: this comment refers to
the distribution of UV flux at $\sim$2200 \AA, which is not the same
as the distribution of Lyman continuum radiation.}

Nevertheless, Figure 1 reveals important activity near the star around
the time of $\eta$ Car's most recent spectroscopic event.  In
particular, the bottom row of images in Figure 1 (panels $d$, $e$, and
$f$) shows pronounced asymmetry that changes with time.  In 2003 Jul
(before the event because of light travel time), we see excess UV
emission toward the SE, and a shadow toward the NW.  Then in 2003 Sep
(Fig.\ 1$e$), there is a UV deficit everywhere (as the ejecta are now
seeing the event itself, when the hot companon is presumably buried in
the primary's optically-thick wind and the far-UV emission is
thermalized).  Later in 2003 Nov, the UV brightness distribution is
the opposite of that just before the event in 2003 Jul -- i.e., now it
is bright toward the NW and shadowed toward the SE.  These distinct
and systematic changes in the observed asymmetry suggest strongly that
$\eta$ Car is indeed a binary system, including a hot companion and a
primary with a very dense wind that blocks UV radiation at scales
larger than the orbital radius at periastron.

Pre-event images in the top row of Figure 1 show little variability,
except for some fading toward the NE of the star in 2003 Jun.  This is
consistent with any model of the event -- in a single star model the
shell ejection has not yet occurred, and in a binary model the
companion is still far enough away from the primary (about 8 AU;
Damineli et al.\ 1997) that its UV radiation is able to escape across
a wide sold angle.

Figure 1 is the first demonstration of severe geometric asymmetry that
changes rapidly around the time of an event, but it is not the first
sign of variable UV flux reaching the ejecta during the 5.5 yr cycle.
{\it HST}/WFPC2 images also showed changes in extended structure on
longer timescales during the 5.5 yr cycle (Smith et al.\ 2000),
continued monitoring at radio wavelengths has shown dramatic
structural changes in the ejecta (Duncan et al.\ 1995, 1997; Duncan \&
White 2003), and near-IR images showed structural changes similar to
those seen in the radio (Smith \& Gehrz 2000). Variability in the
radio and IR traces free-free emission from ionized gas, and hence,
the Lyman continuum radiation field, whereas the F220W filter images
trace longer UV wavelengths riddled with Fe~{\sc ii} absorption in the
wind.  Additionally, van Genderen et al.\ (1999) noted changes in the
near-UV photometry associated with events.  All these types of
variability have been interpreted as arising in equatorial gas, and in
Paper I we argued that the UV excess itself is primarily equatorial.
This emission does not signify a true disk (i.e. a strong equatorial
density enhancement), but rather, it probably arises because UV
radiation is able to escape $\eta$ Car's dense wind more easily at
low-latitudes where the optical depth is usually lower (Smith et al.\
2003a).  Our results reported here modify this idea in the sense that
a hot companion also seems to contribute some of the equatorial UV
excess.  A hot companion contributing a fraction of the observed UV
flux is consistent with the composite nature of $\eta$ Car's far-UV
spectrum (Ebbets et al.\ 1997).  If a hot secondary star contributes
substantially to the observed UV flux, then only the shadow knows the
uncontaminated spectrum of the primary star.

\subsection{Selective Illumination}

Damineli et al.\ (1997, 2000) reported Doppler shifts of broad lines
(especially He~{\sc i}) in ground-based spectra that change
systematically with the 5.5 yr cycle.  Damineli et al.\ interpreted
these shifts as orbital reflex motion in an eccentric binary system
with the major axis of the orbit almost along our line of sight.
However, these velocity shifts around the event were not observed in
high spatial resolution spectra obtained with {\it HST} (Davidson et
al.\ 2000), raising doubts about the orbit solution and the binary
hypothesis.

From our images, it is clear that UV radiation escapes to large radii
outside the wind in preferred directions that vary during the event.
It is plausible, then, that this selective UV illumination of the
ejecta may affect observed line profiles seen in ground-based data.
Directional UV illumination would also operate within the more compact
stellar wind enveloping the binary system, but the effect on the
resulting emission-line profiles may be non-intuitive.  This scenario
might help explain why different emission lines show different
Doppler-shift patterns (Davidson et al.\ 2000), since various lines
arise from different radii in the wind.  This would not be as easily
explained if the observed line profile shifts were due to orbital
reflex motion. In summary, we think the observed ``Doppler shifts''
{\it do not trace the motion of either star} in a hypothetical binary
system, but instead represent the {\it selective illumination} of
certain portions of the outflowing stellar wind and ejecta indicated
by the observed geometry in our UV images.

\subsection{Qualitative Constraints on the Orbit}

If we abandon the idea that observed line profile variations of broad
lines are caused by orbital reflex motion, following the discussion in
the previous section, then we are free to investigate possible
alternative orientations for the orbit.  The observed UV variability
gives vital clues to the orientation of the putative orbit if we
assume that UV radiation from the companion star escapes more easily
in the direction away from the primary star.  Analysis of X-ray
emission indicates that the orbit is likely to be highly eccentric,
with $e\simeq$0.8 to 0.9 (Corcoran et al.\ 2001; Ishibashi et al.\
1999; Ishibashi 2001).

From our data, the tightest constraints on the geometry come from the
illumination of the ejecta just before and after the event --- 2003
Jul and 2003 Nov if we account for light travel time.  Since the
secondary's UV radiation escapes preferentially toward the SE before
the event, and since the illuminated wind and ejecta are primarily
equatorial, this implies that the hot companion star is on the far
side of the primary before the event.  Similarly, the hot companion
would be on the near side of the primary after the event when UV
escapes toward the NW.  This is also supported by the variability of
some faint filaments in the blueshifted ``Fan'' region more than
1$\arcsec$ NW of the star (see Paper I), which appear after the event.
The projected directions of ``shadows'' and excess UV emission suggest
that the major axis of the highly-eccentric orbit is not oriented
along our line of sight, but is instead roughly perpendicular to it,
as sketched in Figure 2. Shadows may be caused by the dense wind of
the primary star absorbing UV radiation from the secondary; thus, the
shadows are seen in the direction opposite that of the secondary's
escaping UV.\footnotemark\footnotetext{Of course, this does not
preclude a more complicated scenario where close passage of a
companion at periastron induces a latitude-dependent shell ejection
(Smith et al.\ 2003a).}  This scenario allows us to constrain the
orientation of the orbit based on the required relative positions of
the primary and secondary (Fig.\ 2).  This is obviously a qualitative
conclusion, but follows from a simple geometric rationale.

Using our images to deduce the direction of apastron is less
straightforward, but the location of the hot secondary during the
high-excitation phase between events is probably toward the SW of the
primary star (at position 1 in Fig.\ 2).  This is because shortly
before the event in 2003 Jun (between positions 2 and 3 in Fig.\ 2),
when the hot companion is approaching very close to the primary, the
UV excess is stronger toward the SW (even at large distances from the
star not shown in Fig.\ 1), and we see a shadow developing toward the
NE of the star at this epoch.  This orbital geometry has many
potential consequences for various structures in the extensive nebula
around $\eta$~Car --- especially the NN ``jet'' and the S Condensation
(Smith \& Morse 2004; Walborn 1976).

When combined with constraints from X-ray observations, these
geometric clues from the UV variability will hopefully be useful to
help predict line profile shapes and other types of observable
changes.  It is relevant to note that an orientation of the major axis
roughly perpendicular to our line of sight is consistent with some
interpretations of the X-ray light curve (Ishibashi 2001; Corcoran et
al.\ 2001).  To our knowledge, Ishibashi (2001) first showed that an
orientation perpendicular to our line of sight provided the best fit
to the asymmetry in the X-ray emission before and after the event; at
the time, this suggestion seemed to contradict the favored binary
model with the major axis pointed toward us (Damineli et al.\ 1997,
2000).

Our observations suggest that $\eta$ Car is in fact a binary system
consistent with interpretations of several other datasets, although
combining all data to formulate a consistent model for the events
remains a formidable task.  A single star would need to perform
impressive contortions to account for the observed asymmetry in UV
images.

\acknowledgements  \scriptsize

Support was provided by NASA through grant HF-01166.01A from the Space
Telescope Science Institute, which is operated by the Association of
Universities for Research in Astronomy, Inc., under NASA contract
NAS~5-26555, and through STIS GTO funding.

\begin{deluxetable}{llll}
\tighten
\tablewidth{0pt}
\tablecaption{Observation Log}
\tablehead{
\colhead{\ } &\colhead{Date} &\colhead{MJD} &\colhead{Phase} }
\startdata
a &2002 Oct 14 &52561.1 &0.906  \\
b &2003 Feb 12 &52682.6 &0.922  \\
c &2003 Jun 13 &52803.1 &0.982  \\
d &2003 Jul 20 &52840.2 &0.000  \\
e &2003 Sep 13 &52896.0 &0.028  \\
f &2003 Nov 14 &52957.8 &0.058  \\
\enddata
\end{deluxetable}

\begin{figure}
\epsscale{0.88}
\plotone{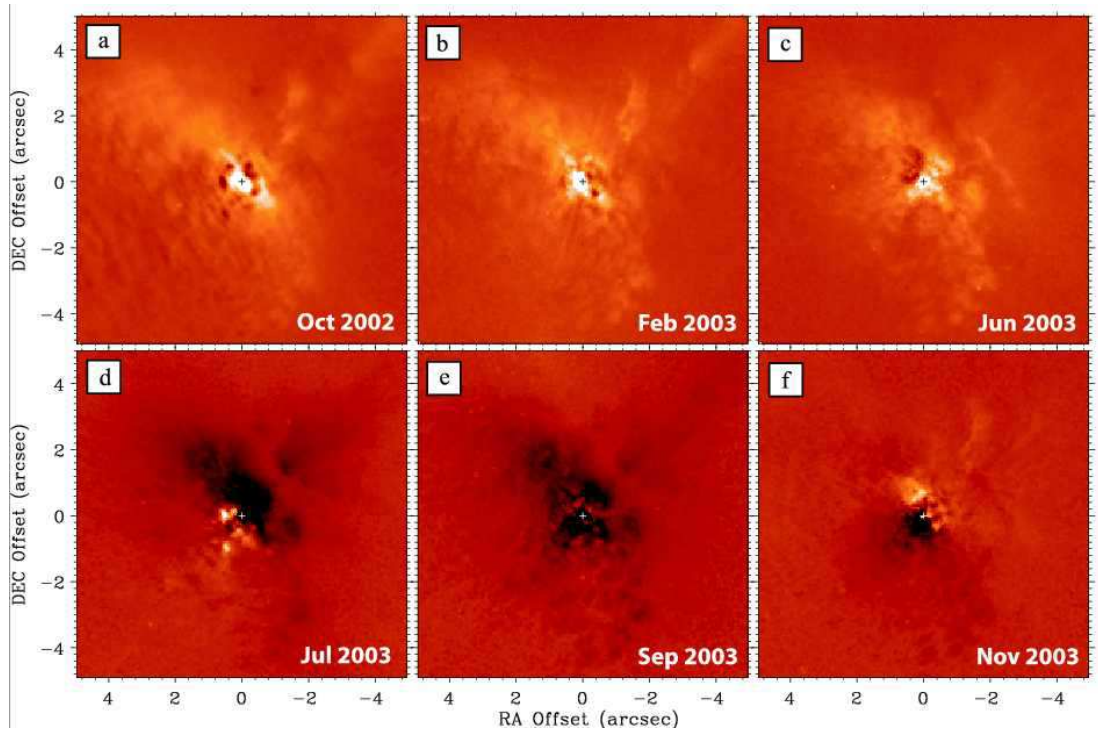}
\caption{Variability in the Purple Haze around $\eta$ Car shown with
difference images in the F220W filter.  Each panel is the image at the
indicated epoch with the average of all epochs subtracted.  Bright
regions are brighter than the $\sim$1 yr average flux, and dark
regions are fainter than average.  The subtraction residual range
shown is roughly $\pm$10$^{-12}$ erg s$^{-1}$ cm$^{-2}$ \AA$^{-1}$
arcsec$^{-2}$.}
\end{figure}

\begin{figure}
\epsscale{0.8}
\plotone{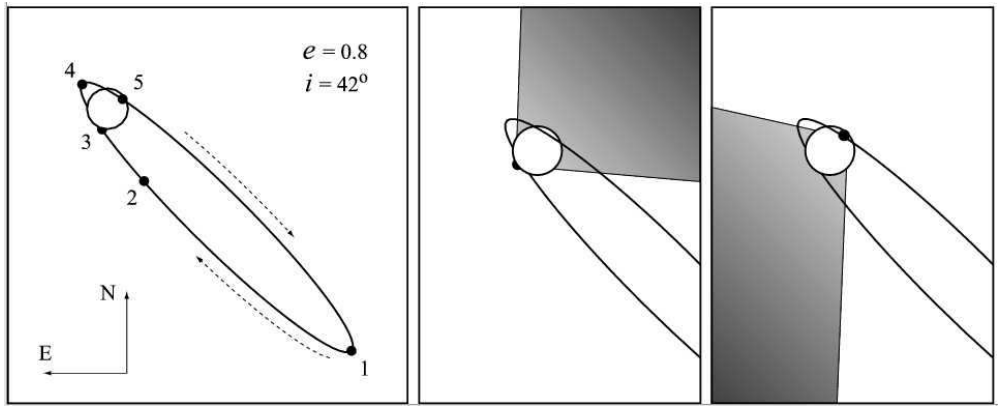}
\caption{The left panel shows a sketch of one possible orbit
orientation of the hot secondary relative to the more massive primary
seen projected on the sky (north up, east to the left).  Filled black
circles represent the secondary at the following times during the
orbit: 1) apastron, 2) a few months before the event in Figs.\ 1$a$,
1$b$, and 1$c$, 3) immediately before the event, as seen by the ejecta
in 2003 Jul (Fig.\ 1$d$) because of light travel time, 4) periastron
(seen by ejecta in 2003 Sep; Fig.\ 1$e$), 5) after the event, as seen
by ejecta in 2003 Nov (Fig.\ 1$f$).  Dashed arrows indicate the
direction of motion. The ``shadow'' projected by UV absorption in the
primary's wind is shown before (corresponding to Fig.\ 1$d$) and after
(Fig.\ 1$f$) the event in the middle and right panels, respectively.}
\end{figure}

\end{document}